\documentstyle{article}

\def\inbar{\vrule height1.5ex width.4pt depth0pt}
\def\C{\relax\,\hbox{$\inbar\kern-.3em{\rm C}$}}

\def\R {{\rm I\!R}}

\def\H {{\rm I\!H}}

\def\beq{\begin{equation}}
\def\eeq{\end{equation}}
\def\bea{\begin{eqnarray}}
\def\eea{\end{eqnarray}}

\def\MM{{  M_{2\times 2}(\C)}}
\def\MMM{{ M_{3\times 3}(\C)}}

\begin{document}

\title{
On variants of the $\C \oplus \MM \oplus \MMM$ NCG model of
elementary particles}
\author{Alejandro Rivero \thanks{Departamento de F\'{\i}sica Teorica,
Universidad de Zaragoza, {\tt rivero@sol.unizar.es} }}
\maketitle

\begin{abstract}
We investigate restrictions to be
imposed in the NCG $C + M_2 +M_3$ model to make it to
fit with phenomenological data. Under strong conditions
over the NCG field a leptophobic $Z'$ boson is got.
\end{abstract}

Recent work \cite{mars} shows that naive application
of Connes's scheme\cite{c.real} to the $\C \oplus \MM \oplus \MMM$ 
algebra drives to a model of elementary particles which
has not an easy phenomenological
fit, nor a trivial method to remove anomalies.

In this short letter, we point out that results can be best fitted 
if we take into account the difference between quark and leptonic
sectors. In Connes' work \cite{c.book}, the bialgebra $\C\oplus\MM ,
 \C\oplus \MMM$ is rejected because we need to get the
 quark yukawa couplings of the standard
model, and such condition is automatically achieved if we take the
bialgebra to be $\C\oplus \H,\C\oplus \MMM$. But no restriction
 was really needed for the
lepton sector. So our path of search can start
 from  $\C\oplus \MM, \C \oplus \MMM$ and
look for conditions restricting the action on quarks to
 be {\em quaternionik}.

Indeed, this can be the case if we demand the fields to
show some kind of
``Independence between actions by $\C$ and by $\MM$''

Plan of letter is as follows: First we look
 some justifications for shrinking the
$\MM$ action to be $\sim \H$ over the quarks. We develop the
calculation in the bimodule formalism, where it is simpler to separate
quark and leptons. Then unimodularity conditions are applied and we
examine the resulting fields, then relating it to the phenomenological
ones. Finally, we conclude with some comments about where to extend
this toy model towards.

Remember that a field $A$ is a first order 
operator $A=\sum a[D,a']$ which is self-adjoint
under the $*$-involution (see \cite{c.book,gv.yellow,cordelia} for
details). For models of type $\cal A= C(M) \otimes A_F$, i.e, an algebra
of continuous functions times a finite matrix algebra, this operator
decomposes in a term due to the $\cal A_F$ and other coming from the
one of continuous functions. The $*$ involution acts as adjunction in the 
finite part and anti-adjuction in the continuous one. 

The finite term of $A$ for the quark part is:
\beq
\pi_q=\sum \pmatrix{
   0&0& m^+_d \lambda (\alpha'-\lambda') & m^+_d \lambda \beta' \cr
   0&0& -m^+_u \bar\lambda \bar\beta' 
                    & m^+_u \bar\lambda(\bar\alpha'-\bar\lambda') \cr
   (\alpha(\lambda'-\alpha')+\beta\bar\beta')m_d 
       &(-\alpha\beta'+\beta(\bar\lambda'-\alpha'))m_u  &0&0\cr
   (-\bar\beta(\lambda'-\alpha')+\bar\alpha\bar\beta')m_d
       &(\bar\beta \beta'+\bar\alpha(\bar\lambda'-\bar\alpha'))m_u &0 &0 } 
\eeq
Where $x, \bar x$ were conjugate complex numbers in the $\C\oplus \H$
model, but now they are independent(We notate the complex 
conjugate as $x^+$).
For the lepton part, the operator is:
\beq
\pi_l=\sum \pmatrix{
    0 & m_e^+ \lambda(\alpha'-\lambda') & m_e^+ \lambda\beta' \cr
   (\alpha(\lambda'-\alpha')+\beta\bar\beta')m_e &0&0\cr
   (-\bar\beta(\lambda'-\alpha')+\bar\alpha\bar\beta')m_e &0&0}
\eeq
  
Now, $A=A^*$ implies two restrictions in both parts; namely:
\bea
\label{adj.diag}
\sum \lambda(\alpha'-\lambda')&=&
       \sum( \alpha(\lambda' - \alpha') + \beta \bar\beta')^+
\\
\label{adj.anti}
\sum \lambda \beta'&=&\sum(-\bar\beta(\lambda'-\alpha')+\bar\alpha\bar\beta')^+
\eea
and the quark part has two additional conditions (which do not apply to
leptons due to the absence of massive neutrino):
\bea
\label{extra.diag}
\sum\bar\lambda(\bar\alpha'-\bar\lambda')&=&
 \sum(\bar\beta\beta'+\bar\alpha(\bar\lambda'-\bar\alpha'))^+
\\
\label{extra.anti}
\sum\bar\lambda\bar\beta'&
  =&\sum(\alpha\beta'-\beta(\bar\lambda'-\bar\alpha'))^+
\eea

Note that if we take the algebra of quaternions, the two last equations
are simply conjugates of the two former.
To clarify calculation, let's define variables that tell us
how much the $M_2$ elements differ from being quaternions:
\bea
\mu&=\bar\beta-\beta^+ \\
\nu&=\bar\alpha-\alpha^+
\eea

With this notation, let us subtract (\ref{extra.diag}) and (\ref{extra.anti}) 
from -the conjugates of- (\ref{adj.diag}),(\ref{adj.anti}) respectively.
We get the relation
\bea
\label{res.diag}
\sum\lambda^+\nu' - \nu^+\lambda'=&& \sum  \mu^+\beta'^+ -(\alpha+\nu^+)\nu'^+ 
                   -\nu^+\alpha' - \beta\mu'
\\
\label{res.anti}
\sum \lambda^+\mu'- \mu\lambda'=&& \sum  \beta^+\nu'^+   - \mu\alpha'
                  -(\alpha^+ +\nu)\mu'  - \nu\beta'^+  
\eea

Now we examine the continuous part, which is the one giving the gauge bosons.
The part coming from $\C$ is $\Lambda=\sum \lambda d\lambda'$, while
the term asociated to $\MM$ has the form
\beq
Q= \sum \left(\pmatrix{\alpha&\beta\cr -\beta^+&\alpha^+}
               + \pmatrix{0&0\cr -\mu & \nu}\right)
        \left(\pmatrix{d\alpha'&d\beta'\cr -d\beta'^+&d\alpha'^+}
          + \pmatrix{0&0\cr -d\mu'&d\nu'}\right)
\eeq

Over this, the condition $A=A^*$ asks $Q$ to be anti-selfadjoint which
implies the following two conditions:
\beq
\label{dif.diag}
\sum
-\beta^+d\mu'^+ +\mu d\beta' -\nu d\alpha'^+ -(\alpha^+ +\nu) d\nu' =
\sum
 \beta d\mu' -\mu'^+d\beta'^+ +\nu^+d\alpha' +(\alpha + \nu^+) d\nu'^+
\eeq
\beq
\label{dif.anti}
\sum
-\beta^+d\nu'^+ = \sum -\mu d\alpha' -\nu d\beta'^+ 
                 -(\alpha^+ + \nu) d\mu'
\eeq

Now we can use 
 (\ref{res.diag},\ref{res.anti}) over the two
last equations to obtain
\footnote{Additional restrictions coming from the non-emptyness of
the kernel of $\pi(\Omega A)^1$ only imply a decrease of freedom
in the RHS of eq. (\ref{dep.diag},\ref{dep.anti}) and do not change the
conclusion}
\bea
\label{dep.diag}
\sum d(-\nu \lambda'^+ +\lambda \nu'^+)-
   d(\nu^+ \lambda'-\lambda^+ \nu') = 
   \sum F(d\alpha,d\beta,d\mu,d\nu,\alpha',\beta',\mu',\nu')&&
\\
\label{dep.anti}
\sum d(\lambda^+ \mu' - \mu \lambda') =
   \sum G(d\alpha,d\beta,d\mu,d\nu,\alpha',\beta',\mu',\nu')&&
\eea

So we can get additional relations between the $\C$ and
$\MM$ algebras, which were tautologies in the $\C \oplus \H$ case.
It's unclear for us if such restrictions have real relevance after
summation. If they had,
 as they enter through the non-cuaternionic part
of $M_2$,  we
would choose $\mu=\nu=\mu'=\nu'=0$ to avoid them. 

Anyway, if we assume directly such restriction $\MM \to \H$ in
 the quark sector, the representation 
of $A$ results in a continuous part:
\beq
\label{gauge.q}
\pi_q(\Lambda,V_0)=\pmatrix{\Lambda & & & \cr
                                  & \bar \Lambda & & \cr
                                  & & V_0 },
\Lambda \in i\R, V_0 \in \H
\eeq 

Per contra, as (\ref{res.diag},\ref{res.anti}) do not 
appear in the lepton side, we choose do not restrict it, and
the corresponding term is given by:
\beq
\label{gauge.l}
\pi_l(\Lambda,V_0,B)=\pmatrix{\Lambda &  \cr
                            &    V_0+ B },
V_0 \in \H; \ \Lambda,B \in i\R
\eeq

With this, the action of the bimodule for the hilbert space
${\cal H}= h_l \oplus (h_q \otimes \C^3)$ can be writen 
(with $K\in M_3, U,\Lambda,B \in i\R, V_0 \in \H$) as:
\beq
\pi((\Lambda,V_0,B),(U,K))= (\pi_l(\Lambda,V_0,B)+U) \oplus
                           (\pi_q(\Lambda,V_0) \otimes K)
\eeq

Now, we apply unimodularity conditions in the old style \cite{c.book,gv.yellow}
\beq
N_g(\Lambda+U) + 2 N_g \mbox{Tr} K =0
\eeq
\beq
2N_g B + 2 N_g U + 2 N_g \mbox{Tr} K=0
\eeq
$N_g$ being the number of generations.

From this, we got the relationships
\beq
\Lambda=U+2B
\eeq
and
\beq
(U+B)+ {\mbox Tr} K =0
\eeq

Rewriting 
\beq
\label{redef.a}
A_0 = U + B
\eeq  
\beq 
\label{redef.k}
K_0=K+ {1 \over 3} A_0
\eeq
 we finally get \footnote{ with basis
$\pmatrix{e^R & d^R & u^R & (e,\nu)^L & (d,u)^L}$}
\beq
\pi(A_0,V,B,K_0)=\pmatrix{2 A_0 & & & & &  \cr
                    & {2\over 3} A_0+K_0-B  &&&           \cr
                    &&  -{4\over 3} A_0+K_0+B  &&           \cr
                    &&&  A_0+V_0  &           \cr
                    &&&&  -{1\over 3} A_0+V_0+ K_0             \cr   }
\eeq
where $A_0$ coincides with the $U(1)$ field of standard model, with the correct
hyperchargues, $V_0$ is the $SU(2)$ electroweak field, $K_0$ is the
$SU(3)$ color field
and $B$ is a new boson field coupling only to quarks.


The resulting model is not
anomaly-free. But we are not going to address anomalies (coming from 
the mixed $U(1)_{A_0}-U(1)_B$ triangles) here. Simply note that 
no cancellation mechanism seems available in this small framework.

Note that $B$ is leptophobic, as required by recent studies
\cite{chiap} on new electroweak physics. Moreover, we can suppose
that its coupling constant, $g_2$, is the 
same that the one of the $SU(2)$ electroweak
group, as both fields come from the $U(2)$ field associated to the
$\MM$ algebra.
 
New axial and vector currents associated to this, say, $Z'$ field,
are zero in the lepton sector. For quarks, we get
\bea
t_V=+g_2/4 & t_A=-g_2/4
\eea
on quarks u,c,t, and same with opposed signs for d,s,b:

Doing the quotient by the $Z_0$ currents, we get the 
numbers:
\begin{itemize}
\item For leptons
\bea
l_V=0 & l_A=0
\eea
\item For quarks u,c,t:
\bea
t_V= +{1\over 2} {\cos^2 \theta_w 
  \over \sin \theta_w (\frac12-\frac43 \sin^2 \theta_w)}\approx 4.16
& t_A=-  cos\theta_W \approx -0.87
\eea
\item For quarks d,s,b:
\bea
b_V=- \frac12 {\cos^2 \theta_W \over
         \sin \theta_W (\frac12-\frac23 \sin^2 \theta_W)}\approx 2.32
& b_A=  cos\theta_W \approx 0.87
\eea
\end{itemize}
which we can compare with the experimental fit \cite{chiap}
\def\mz{{{ M_{Z'}\over 1\mbox{TeV} }}}
\bea
\pm l_V =-2.25\pm 6.25\mz  && \pm l_A =-0.2\pm 0.5\mz \\
\pm b_V =-3.45\pm 20.72\mz && \pm b_A =+4.58\pm 9.84\mz \\
\pm c_V =-6.94\pm 26.6\mz  && \pm c_A =-7.88\pm 8.46\mz
\eea
got from LEP results. We see that the new interaction 
could fit with the phenomenology, but present limits on 
$Z'$ mass \cite{agenda} suggest a slightly higher or more 
sophisticated coupling.


To summarize, we draw three conclusions:
\begin{itemize}
\item 
It seems valid, at least operationally,  to restrict the representation
of the fields in the $\C \oplus \MM$ algebra to be the ones of
$\C \oplus \H$
in the quark subspace.
\item
From a representation of this kind, $\C \oplus \H$ over quarks,
$\C \oplus \MM$ on leptons, both the standard and the "bizarre"
\cite{bizarre} distribution of hyperchargues appear.
\item
The new model continues being compatible with the
experimental data.
\end{itemize}

Anomaly conditions have not been examined here. Same with the Higgs, which in
this setup takes a delicate shape; we need to look how many higgses
we have, and which one has the correct quantum numbers to confer mass
to the new field. 

Such questions are delicate to establish in the model,
but the main goal of this letter is only to point a possibility.
In fact, we are doing in some sense a leap of faith when jumping
from equations (\ref{dep.diag}-\ref{dep.anti}) to 
result (\ref{gauge.q}), as we assume that
such equations have different implications that the one we can get from
(\ref{adj.diag}) and the anti-selfadjointness of the diagonal part of the
quaternion. 

It rests to do some small comments about possible developments.
Lets remark again that this presentation is not 
a definitive one. Serius model building will be done actually in the
mood of \cite{c.real,c.cov} to incorporate the Tomita operator.
As pointed in
\cite{c.cov}, the final model would be clearly related to $SU_q(2) \otimes
SU_q(2)$, not to the single $SU_q(2)$ as it is said to happen here. 
And perhaps the last
word on anomaly cancellation would be say in the framework of a
completely unified theory (in the shape of \cite{cc}?), where
 mechanisms as Green-Swartz
cancellation\cite{ibanez.gs,gonzalez.gs},
horizontal symmetries \cite{ibanez.mass}, etcetera, could be available.


Author want ack. input and discussions with F. Falceto, J.L. Cortes
and the UCM-UCR NCG team. Specifically, CP Martin must be acknowledged by
showing us the anomalous triangles of the model.
In addition, JL Cortes must be acknowledged by volunteer to make the digest
of recent accelerator results and explain them in the DFTUZ seminar.

Housing of Complutense Univ, DFTUZ, and Spanish Navy have
been used in the research period.
This work has been delayed by non voluntary unwilling military duties.

\end{document}